\documentstyle[psfig,aps,prl,epsf,floats]{revtex}
\begin{document}
\twocolumn[\hsize\textwidth\columnwidth\hsize\csname %
@twocolumnfalse\endcsname
%\draft
\title{Theoretical study of the optical conductivity of $\alpha'$-NaV$_{2}$O$_{5}$}
\author{Mario Cuoco$^{a,b}$, Peter Horsch$^{a}$, and Frank Mack$^{a}$ }
\address{$^a$ Max-Planck-Institut f\"ur
    Festk\"orperforschung,Heisenbergstr. 1,D-70569 Stuttgart (Germany)}
\address{ $^b$ I.N.F.M. di Salerno,
Dip. Scienze Fisiche ''E.R. Caianiello'',
I-84081 Baronissi Salerno (Italy)}
%\date{\empty}
\vspace{0.2in}
\maketitle
\begin{abstract}
Using finite temperature diagonalization techniques it is shown 
that the quarter-filled $t$-$J$-$V$ model on a trellis lattice structure
provides a quantitative explanation of the
highly anisotropic optical conductivity of the $\alpha'$-NaV$_{2}$O$_{5}$ compound.
The combined effects of the short-range Coulomb interaction and 
valence fluctuations of V-ions determine the main absorption and the fundamental gap.
Inter-ladder hopping is necessary for the explanation of 
the anomalous in-gap absorption and generation of spectral weight at high energy.
The role of valence fluctuations is explained in terms of the  domain wall excitations
of an anisotropic 2D Ising model in a transverse field close to criticality.
\end{abstract}
 ]

\narrowtext
The physical properties of low dimensional quantum systems 
received considerable attention from both theoretical and experimental sides
because of their unconventional spin and charge excitation spectra\cite{rice}.
In this respect, the vanadate $\alpha'$-NaV$_{2}$O$_{5}$, initially 
identified as an inorganic spin-Peierls (SP)\cite{isobe} compound, 
similar to CuGeO$_{3}$, has 
attracted great interest as a ladder system at quarter filling.
%with an intriguing interplay between spin and charge degrees of freedom.
The original X-ray structure analysis\cite{carpy} was pointing in favor of a
non-centrosymmetric structure, implying two inequivalent vanadium
sites. Hence,
it was concluded that  magnetic spin-1/2 V$^{4+}$ ions were forming
one-dimensional (1D) chains well separated by nonmagnetic V$^{5+}$ chains in
the $(a,b)$-plane. Based on this view the insulating properties appear natural, 
and various theoretical studies were performed using dimerized
Heisenberg chains\cite{augier} or spin-phonon models\cite{poiblanc},
which provided satisfactory results, e.g. for the susceptibility.

However, recent structure studies\cite{schnering,smolinski,meetsma} at room temperature 
showed that the V-ions are all equivalent with valence 4.5, which
came as a surprise in view of the insulating properties of this substance. 
This puzzle was resolved by realizing that the molecular orbital state on
a rung occupied by one electron is a key element of the electronic structure.
In combination with the large local repulsion $U$, which confines the formal
charges of V-ions to $+4$ and $+5$, respectively, the insulating behavior
and the quasi-1D magnetic properties could be 
explained\cite{smolinski,horsch}.

Surprisingly
recent NMR-studies\cite{nmr} revealed the appearance of two different V-sites 
below the structural transition at $T_c=34$ K\cite{isobe}, 
which clearly indicates that the transition
is not of spin-Peierls type but rather driven by charge ordering. 
There have been several further experiments showing anomalous features
compared to the usual spin-Peierls transition. The ratio between 
spin gap and transition temperature 
$2\Delta/k_{B} T_{c}\sim 6.5$ \cite{fujii,neutron} is much larger than
the BCS mean-field value 3.52.
The changes at $T_c$ of the dielectric constant\cite{smirnov}
and the thermal conductivity\cite{vasilev} differ significantly from the 
case of CuGeO$_{3}$ .
Moreover the magnetic field dependence $\Delta T_c \propto H^2$ is only about 20\%
of what is expected for a spin-Peierls transition\cite{schnelle}.
%Besides, the Raman scattering experiments\cite{milan,lemmens} show new modes
%in the low temperature phase whose origin is not yet completely understood.

Subsequently several theoretical papers analyzed the role of nearest and further
neighbor Coulomb interaction, proposing 
zig-zag\cite{fukuyama,khomskii} or linear\cite{fulde} charge order in the ground state.
The nature of the excitation spectra of such a
charge ordered singlet ground state is an open question\cite{fukuyama}.
A test of these concepts could emerge from the very distinct
experimental optical conductivity data\cite{damascelli,optsmirn}.
These experiments show quite different optical interband transitions for 
$a$- and $b$-polarization, i.e., reflecting the anisotropic structure of VO-layers.
In addition to the larger spectral weight observed for $a$-polarization, the 
$a$-spectra show weak low energy in-gap absorption\cite{damascelli}.

\begin{figure}
      \centerline{\hbox{
      \epsfxsize=6cm
      \epsffile{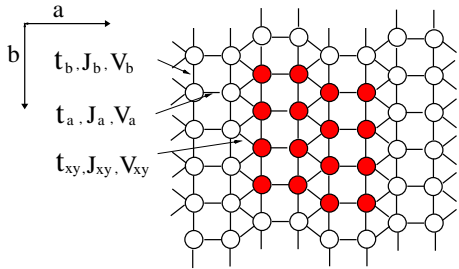}}}
\smallskip
\caption {
Schematic structure of the $\alpha'$-NaV$_{2}$O$_{5}$ $(a,b)$-planes where circle
stands for a vanadium site. The black sites define the size of the cluster used
in the calculation.} 
\label{fig1}
\end{figure}

We study the optical response assuming that the dynamics of electrons is
described by an extended $t$-$J$-$V$ model\cite{riera,ohta} including intra- and inter-rung
Coulomb repulsion for a 2D system with trellis topology shown in Fig. \ref{fig1}.
While in a previous numerical study\cite{ohta} zig-zag order was imposed, our results
show that the excitation spectra are not only determined  by the interladder
coupling but also by strong intra-ladder valence fluctuations.
Our study suggests that  $\alpha'$-NaV$_{2}$O$_{5}$ is close to a quantum critical
point.
The  $t$-$J$-$V$ model is given by:
\begin{eqnarray}
H &=& 
\sum_{\langle i,j\rangle,\sigma} t_{ij} (\tilde{c}^{\dagger}_{i,\sigma}
\tilde{c}_{j,\sigma}+H.c.)\nonumber\\
&+&\sum_{\langle i,j\rangle} J_{ij} ({\bf S} _{i} \cdot {\bf S}_{j}
-\frac{1}{4} n_{i} n_{j})
+\sum_{\langle i,j\rangle} V_{ij}~n_{i} n_{j}
\label{eq1}
\end{eqnarray}
where $\tilde{c}^{\dagger}_{i,\sigma}={c}^{\dagger}_{i,\sigma}
  (1-n_{i,-\sigma})$ are constrained electron creation operators,
$n_{i}=\sum_{\sigma}\tilde{c}^{\dagger}_{i,\sigma}\tilde{c}_{i,\sigma} $, 
and ${\bf S} _{i}$ is the spin-$\frac{1}{2}$ operator.
The sums are over all bonds $\langle i,j\rangle$ between neighbors on the trellis lattice  
and spin $\sigma=\uparrow,\downarrow$. 
The various hopping parameters $t_{ij}$, intersite Coulomb interactions $V_{ij}$, 
and exchange interactions $J_{ij}$
between vanadium neighbors are defined in Fig. \ref{fig1}.
The exchange term is parametrized as $J_{ij}=4 t^2_{ij}/U$, where $U$ is the
corresponding on-site Hubbard interaction and we assume that $U=4.0$ eV. 
The calculation of $\sigma(\omega)$ has been performed by means of a 
generalization of the 
Lanczos technique for finite temperature (FTLM)\cite{prelovsek}.
In this approach the trace of the
thermodynamic expectation value is evaluated by Monte-Carlo sampling,
while for the evaluation of the relevant matrix elements the Lanczos
method is used. 
Since the system is
in an insulating state, one can restrict the calculation of 
$\sigma(\omega)$ to the finite
frequency response given by the Kubo formula
\begin{eqnarray}
\sigma_{m}(\omega)=\frac{1-e^{-\beta \omega}}{\omega} 
Re \int_0^{\infty} d\tau e^{i \omega \tau} \langle j_m(\tau)j_m \rangle ,
\end{eqnarray} 
where $\beta=1/k_{B}T$ and $j_{m}$ with $m (=a,b)$ denoting a component of the
current operator
\begin{equation}
j_{m}=i \sum_{\langle i,j\rangle,\sigma} t_{ij} \cdot R^{ij}_{m}
(\tilde{c}^{\dagger}_{i,\sigma} \tilde{c}_{j,\sigma}-H.c.).
\end{equation}
\noindent
Here $R^{ij}_{m}$ is the $m$-th component of the vector connecting neighboring 
sites $i$ and $j$. 

\begin{figure}
\centerline{\psfig{figure=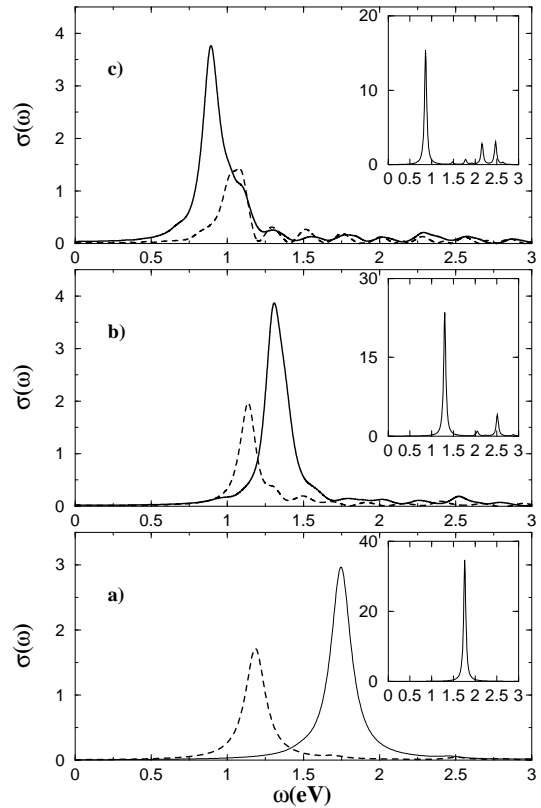,width=7cm}}
\narrowtext
\smallskip
\caption
{Optical conductivity $\sigma_a(\omega)$ (full line) and $\sigma_b(\omega)$  
(dashed line) for a trellis lattice as shown in Fig.1. 
The parameters used are $V_a=V_b=$0.8, $t_a=$0.4,
  $t_b$=0.2, $t_{xy}$=0. eV at a temperature T=300 K. From a) to c) 
$V_{xy}=0.,0.8,1.2$ eV, respectively. Inset: Transverse response 
$\langle T^{y}_{-\bf q} T^{y}_{\bf q}\rangle_{\omega}$ for
$a$-polarization at ${\bf q}=(0,0)$ for the pseudospin model.}
\label{fig2}
\end{figure}

\begin{figure}
\centerline{\psfig{figure=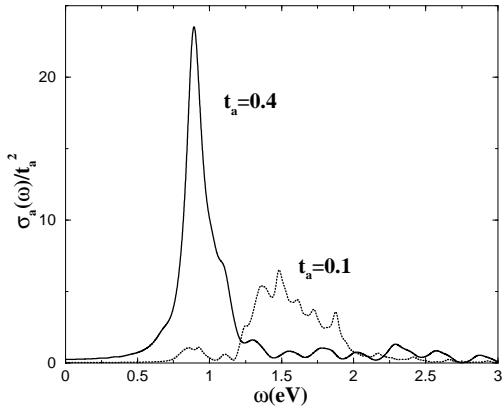,width=7cm}}
\narrowtext
\smallskip 
\caption 
{Comparison of normalized optical conductivity $\sigma_a(\omega)/t_a^2$ for  
$t_a=0.1$ and $0.4$ eV which documents the effect of valence fluctuations; 
other parameters as for  Fig. \ref{fig2}c.}
\label{fig3}
\end{figure}

First we consider the optical spectra for a system of  ladders
coupled only by the interaction $V_{xy}$, i.e., putting $t_{xy}=0$.
In Fig. \ref{fig2} we compare results for spectra with polarization along $a$ and $b$
direction, respectively, for a cluster formed by two ladders whose size is
shown in Fig. \ref{fig1}. Periodic boundary conditions are chosen.
We use $t_a=0.4$ and $t_b=0.2$ eV \cite{smolinski,horsch}.
For the Coulomb repulsion we consider 
the symmetric condition $V_a=V_b=V$ \cite{fukuyama}
with $V=0.8$ eV and vary $V_{xy}$.
The $a$-spectra show a strong downward shift when increasing the
value of $V_{xy}$, whereas the $b$-spectra hardly change. This behavior reflects
the different nature of excited states relevant for $a$- and $b$-polarization.
Polarization along $b$ connects to excitations 
with two and zero electrons per rung, respectively.
The dynamics of these `doublons' and `holons'
is not much influenced by valence fluctuations, since these rungs
are fully occupied or empty, and also not by  
$V_{xy}$, since the interaction is frustrated (Fig. 1).
The nontrivial $V_{xy}$ dependence of the $a$-spectra will be discussed below.
Here we only emphasize that 
the  $a$-spectra are sensitive to the ratio $2 t_a/V_b$.
Indeed, as one can see in Fig. \ref{fig3}, if valence fluctuations are weak 
(small $t_a$), the main weight in the
$a$-spectrum is at energy $\sim 2 V_{b}$, as  expected for the case of
decoupled ladders (see Fig. \ref{fig2}a), but different from the result
for larger $t_a$.

\begin{figure}
\centerline{\psfig{figure=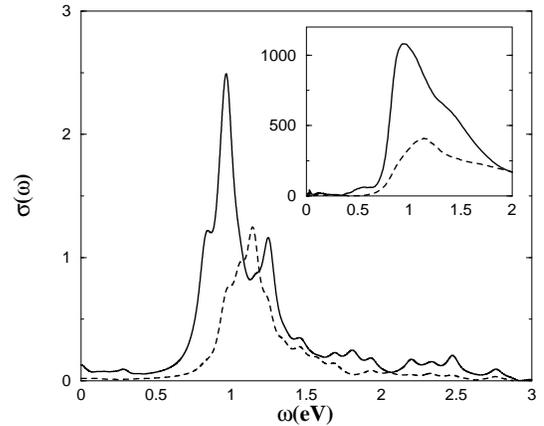,width=7cm}}
\narrowtext
\smallskip
\caption
{Optical spectra for $a$-(full line) and $b$-(dashed line) polarization: the
  parameters are the same of Fig. \ref{fig2}c but for finite hopping $t_{xy}$=0.15, and
  $V_{xy}=0.9$ eV at T=300 K. The inset shows experimental data by Damascelli
  {\it et al.}$\protect\cite{damascelli}$.
  The ratio $r=I_a/I_b$ 
  of $a$- and $b$-intensities integrated from 0.6 to 2 eV is
  $r_{theor}=1.9$ and $r_{exp}=2.2$, respectively.  }
\label{fig4}
\end{figure}

The numerical data in Fig. \ref{fig2}c for $t_{xy}=0$ contains already some of the 
characteristic features of the optical experiments, such as the relative peak positions
of the spectra with $a$ and $b$ polarization and also their different spectral
weight. 
Next we  consider the role of the hopping $t_{xy}$ along the zig-zag
chains of edge sharing VO$_5$ pyramids.
In Fig. \ref{fig4} results are shown for
$t_a=0.4,t_b=0.2,t_{xy}=0.15$ and $V_a=V_b=0.8, V_{xy}=0.9$ eV, where 
the $t_{xy}$ value was optimized to have
the best agreement with the experiments given in the inset\cite{damascelli}.  
One important effect of finite $t_{xy}$ is the generation of additional spectral weight 
on the high energy side of the spectrum. We attribute this to the incoherent motion
of excitations along the $a$-direction due to the finite $t_{xy}$.
Compared to Fig. \ref{fig2}c the $a$-spectrum shows a prepeak at 0.8 eV and a peak at 
1.3 eV due to inter-ladder excitations, which falls in the range of the  shoulder in the 
experimental spectra.
There is also a weak low energy absorption below 0.3 eV, reminiscent
of the `charged magnon' absorption observed in infrared optics in the range
$\omega \sim$ 0.01-0.15 eV\cite{damascelli}.
We believe that these are charge excitations, which are  activated by 
spin-charge  and inter-ladder coupling, 
as they disappear in the spinless fermion case
or when the hopping between the ladders is switched off (Fig. \ref{fig2}c).

For $b$-polarization the maximum of the absorption is at 1.15 eV, i.e., slightly higher
than the main peak in the $a$-spectrum. Again there is high energy spectral weight,
although less pronounced, which extends up to 2.5 eV.
There is no  evidence of low energy excitations in $\sigma_b(\omega)$,
consistent with their absence  in experiments.

Temperature effects are small up to 500 K, however
there is a decrease of
spectral weight in the a-polarization spectrum for $\omega<0.2$ eV and T$<200$ K. 
For the chosen parameters, the equal-time
density correlation functions indicates a tendency towards a zig-zag pattern,
although the charge distribution itself is uniform.

To get a deeper understanding of the optical properties 
we turn now to a different representation of the low energy
degrees of freedom. For large $U$ and $V_a$
the relevant subspace of one electron per rung can be
represented as eigenstates of spin ${\bf S}$ and pseudospin
${\bf T}$ operators, where the configurations with 
$T^{z}=\pm \frac{1}{2}$ correspond to the left/right position of the electron
within a rung. 
After projecting out the high energy states we obtain
the following Hamiltonian  for a single ladder\cite{khomskii}:
\begin{eqnarray}
H_{ladder}=-2 t_{a}\sum_{r} T^{x}_{r} +(2 V_{b}+\frac{2 t^{2}_{b}}{\Delta})
\sum_{r} T^{z}_r T^{z}_{r+b} \nonumber\\
+\frac{8 t^{2}_{b}}{\Delta} \sum_{r} ({\bf S}_r \cdot {\bf
  S}_{r+b}-\case{1}{4})(T^{+}_r T^{-}_{r+b}+H.c.)
\label{eq2}
\end{eqnarray}
where $r$ indicates a composite site formed by the two vanadium ions in the
rung. The hopping $t_a$ acts like a
transverse field, while the Coulomb interaction between electrons on
neighbor rungs $r$ and $r+b$ in the ladder is like the $zz$ term in the Ising
model.
The third term is due to the exchange  
between electrons in different rungs along the $b$ direction. The value of $\Delta$
scales between
$V_{a}+V_{b}$ and $2 t_a$ depending on which  scale of energy is larger.

The coupling between neighboring ladders yields
\begin{eqnarray}
H_{inter-ladder}=-V_{xy} \sum_{\langle r_1,r_2 \rangle } (T^{z}_{r_1}-\case{1}{2})
(T^{z}_{r_2} + \case{1}{2})\\
+\frac{4 t^{2}_{xy}}{\Delta_1} \sum_{\langle r_1 r_2 \rangle} ({\bf S}_{r_1} \cdot {\bf
  S}_{r_2}-\case{1}{4})(T^{z}_{r_1} T^{z}_{r_2}+\case{1}{4})\nonumber
\end{eqnarray}
with $r_2=r_1+(\frac{a}{2},\pm \frac{b}{2})$ and $\Delta_1=2 V_b+V_a-V_{xy}$.
In this representation $\sigma_a(\omega)$ is given by the
transverse response function 
$\langle T^{y}_{-\bf q} T^{y}_{\bf q}\rangle_{\omega}$
with ${\bf q}=(0,0)$ if $t_{xy}=0$.
The latter is shown in the insets of Fig. \ref{fig2} and has been evaluated
by means of the FTLM
using the same parameters as for the $t$-$J$-$V$ model and
assuming that the spin correlations can be treated as for valence-bond states, i.e., 
$\langle {\bf S}_{r_1} \cdot {\bf S}_{r_2} \rangle =-\frac{3}{4}$.
As one can see the main peak position and its $V_{xy}$ dependence are properly
described. 
The dominant pole structure is consistent with the excitonic nature of the
optical absorption. The spectral broadening in the complete $t$-$J$-$V$ model
can be attributed to the dynamical character of the spin degrees of freedom.

\begin{figure}
\centerline{\psfig{figure=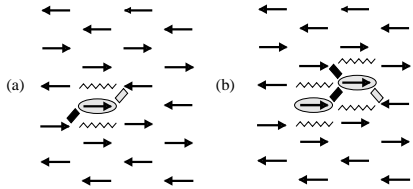,width=6cm}}
\narrowtext
\vspace{5mm}
\caption {\label{fig5}
Configuration with one (a) and two pairs of domain-walls (b). The
pseudospin pointing to the left (right) direction indicate a charge
configuration with the electron on the left (right) position in the rung,
respectively. Black (gray) stripes show the bond where electrons of neighboring
rungs which belong to
different ladders feel the Coulomb repulsion $V_{xy}$ before (after) flipping one 
(a) and two (b) pseudospins. 
}
\end{figure}

The strong $V_{xy}$ dependence of the peak position 
in the $a$-spectrum cannot be understood if $2t_a/V_b$ is small, i.e., when the
ground state has a simple zig-zag charge order pattern, corresponding to a
classical N\'eel order of pseudospins. 
In this case due to the geometry of the inter-ladder bonds
the flipping of a pseudospin, i.e., moving an
electron from left (right) to right (left), does not change the number of
frustrated bonds with energy $\sim V_{xy}$, and the excitation energy is $2V_b$.
This case is sketched in Fig. \ref{fig5}a, which shows a single flipped spin in an 
otherwise N\'eel ordered arrangement. 
By flipping a further pseudospin in the neighboring ladder,
however, e.g. by an optical excitation,
one can reach the configuration of Fig. \ref{fig5}b with one frustrated bond less 
compared to the starting configuration (Fig. \ref{fig5}a), which leads to 
a reduction of the excitation energy ~$V_{xy}$. 
Such processes can be neglected in the  $2t_a/V_b\rightarrow 0$ limit.
However, when the system is close to its critical point,
i.e., $2 t_a/V_b\sim 1$, the ground state is dominated by 
configurations with a high density of domain
walls. In this case most optical excitation processes occur in the neighborhood
of already existing domain walls. 
This implies that near criticality even though the energy of the
ground state is not strongly affected by $V_{xy}$, 
the excitation spectrum undergoes significant changes. The excitation energy 
is lowered by the repulsion $V_{xy}$ as shown in Fig. \ref{fig2}.

Finally, we address the low-energy features appearing in the $a$-spectra
by considering the single chain Ising model in a transverse field,
i.e., given by the terms $\propto t_a$ and $V_b$ in Eq. (4). 
The excitation spectrum of this model is described by the 
domain wall dispersion of the form $\Lambda_{q}=V_b \sqrt{1+h^2+2h \cos(q)}$, 
with $h=2 t_a/V_b$ \cite{lieb,sachdev}. 
There are low-energy charge excitations near $q=\pi$ connected to the
slow dynamics of the domain walls close to the critical point,
which can however not be optically excited.
The optical response for $a$-polarization is determined by the excitation 
of $q\sim 0$ pairs
of domain walls whose energy only slightly deviates from the static charge gap
value $2 V_b$.
Yet this supports the idea that the low-energy excitations in Fig. \ref{fig4} 
can be attributed to these low-energy charge excitations, namely 
optically activated by spin scattering and by the interladder coupling.
{This can be corroborated by a simplified version of Hamiltonian (4) where 
the spins are assumed to be disordered and static, i.e., by introducing
random variables for $\langle {\bf S}_r \cdot {\bf S}_{r+b}\rangle$.}
In contrast to the `charged magnon' scenario\cite{damascelli} 
the energy scale is determined here by
the domain wall motion.

We conclude that  the optical response at room-temperature can be
well described by the quarter-filled $t$-$J$-$V$ model, 
assuming that the system is close to criticality $2t_a \sim V_b$.
Hence, strong valence fluctuations connected to domain wall dynamics, which have not
been considered in previous studies, play a
crucial role for the understanding of the charge ordering transition and the charge
pattern of the ground state in the low temperature phase.
The comparison with the experimental data also indicates the
importance of both the
inter-ladder hopping $t_{xy}$ and the Coulomb repulsion $V_{xy}$.
Finally, we note that the nature of the single particle excitations for the present model
is a completely open issue. 
Spectral distributions obtained recently by 
ARPES measurements are highly anomalous\cite{Kobayashi98}, but have been
interpreted by the charge-spin 
separation of a doped Heisenberg chain. 
It is natural to expect that the domain wall
excitations significantly influence the character 
of the single particle electron Green's function.

It is a pleasure to thank J. van
den Brink, A. Damascelli, M.J. Konstantinovi$\acute{c}$, A. M. Ole\'s and R. Zeyher 
for helpful discussions and comments.

%
%\end{multicols}
\end{document}